# An inversion in the wiring of an intercellular signal: evolution of Wnt signaling in the nematode vulva

Marie-Anne Félix
Institut Jacques Monod, CNRS-University of Paris 6-7,
Tour 43, 2 place Jussieu, 75251 Paris cedex 05, France.

Keywords : Wnt signaling, pathway evolution, nematode vulva, *Caenorhabditis elegans*, *Pristionchus pacificus*

**Summary**
**Signal transduction pathways are largely conserved throughout the animal kingdom. The repertoire of pathways is limited and each pathway is used in different intercellular signaling events during the development of a given animal. For example, Wnt signaling is recruited, sometimes redundantly with other molecular pathways, in four cell specification events during *Caenorhabditis elegans* vulva development, including the activation of vulval differentiation. Strikingly, a recent study finds that Wnts act to *repress* vulval differentiation in the nematode *Pristionchus pacificus* [1], demonstrating evolutionary flexibility in the use of intercellular signaling pathways.**

**Introduction**
Diversification of cells during animal development relies largely on intercellular signaling events. The signals and their transduction pathways are conserved throughout animal taxa, and the repertoire is amazingly narrow: the Notch, Wnt, EGF-Ras, Hedgehog, TGFβ, insulin pathways govern most intercellular signaling events. Each of these pathways is therefore reused multiple times during development. The interpretation of the signal within the receiving cell depends on the preexisting state of the cell (its « competence »), either through the expression of specific paralogues of components of the signal transduction machinery and/or of specific downstream transcription factors.

Most developmental studies concern non-homologous processes in very distant model organisms (such as the *Drosophila* eye, the *C. elegans* vulva or the chicken neural crest) and therefore cannot address the question of the evolutionary dynamics of signaling pathway recruitment for a given developmental process. De novo recruitment of these pathways occurs in the development of evolutionary novelties, such as butterfly wing eyespots [2]. However, it could be largely expected that, once a signaling pathway has been recruited, its use would be rather stable over evolutionary time.

Previous studies on *C. elegans* vulval development had shown that Wnt signaling helps to activate vulval cell fates in *C. elegans* [3-7]. The recent finding by Zheng, Messerschmidt, Jungblut and Sommer [1] that it is used as a repressor of vulval cell fates in the nematode *Pristionchus pacificus* thus comes as a surprise, and adds significantly to our understanding of the evolutionary versatility of signaling pathways.

**Background: evolution of nematode vulva development**
The formation of the *Caenorhabditis elegans* vulva involves multiple intercellular signaling pathways. The vulva develops from a set of precursor cells, the Pn.p cells, which are aligned along the antero-posterior axis in the ventral epidermis. In *C. elegans*, six of these cells, numbered P(3-8).p, are competent to receive signals that specify three fates in a centered pattern (Fig. 1). These signaling events have been studied over the last 25 years through a combination of cell ablations with a laser beam and of genetic analyses [8,9]. The anchor cell of the uterus, located close to P6.p, sends an EGF-like signal that induces the central vulval fate (1°) of P6.p, and may act in a graded manner further away from its source. In addition, lateral signals (Delta-like) sent by P6.p in response to EGF-Ras activation induce the 2° fate and repress the 1° fate in P5.p and P7.p. A negative signaling pathway from the surrounding hyp7 syncytium provides a threshold of activation of vulval fates and thus helps to maintain non-vulval (3°) fates in P3.p, P4.p, and P8.p [10,11]. To each of the three cell fates corresponds a specific cell division pattern. The pattern of fates and divisions is centered around the anchor cell and P6.p: especially, the division patterns of P5.p and P7.p have a mirror-image symmetry; the orientation of P7.p divisions depends on Wnt signaling (Fig. 1, step 3)[12].

    The cell lineage invariance among *C. elegans* individuals extends to other nematodes. Especially, the same twelve Pn.p cells and the same pattern of fates for P5.p, P6.p and P7.p are found in a large group of species, that includes the family Rhabditidae, to which *C. elegans* belongs, and the family Diplogastridae, to which *Pristionchus pacificus* belongs. This precursor cell invariance makes the vulva an attractive system for studying the evolution of a homologous developmental process [13-15]. Cell ablation studies have uncovered ample variations in the requirement for the anchor cell in activating vulval fates, including no requirement in some species [16], partially redundant specification [17,18], and requirement for the gonad or the anchor cell at multiple steps [17,19-22]. Yet the molecular identity of the signaling systems was unknown.

    *Pristionchus pacificus* was chosen by Ralf Sommer et al. as a « satellite » genetic model species for comparisons with *C. elegans*. The original reason for choosing *P. pacificus* was the size of its vulval competence group and its regulation by apoptosis [23] (Fig. 1), rather than the conservation of the vulval fate pattern of P(5-7).p. It comes therefore as a surprise than this conserved spatial pattern relies on very divergent intercellular signals.

**Wnt signaling represses vulval fates in *Pristionchus pacificus***

    Screens for vulva mutants in *P. pacificus* yielded several classes of mutations that affect Pn.p fate specification. One phenotypic class includes seven alleles of the same gene, initially called *ped-7*; they display a reverse orientation of the cell division pattern of P7.p, thereby creating a secondary invagination and a small ventral protrusion in the adult [1]. In addition, these mutations cause a hyperactivation of vulval fates in two experimental contexts: i. after gonad ablation, vulval differentiation is not prevented in the *ped-7* mutant, whereas it is fully abolished in the wild type; ii. in the *ped-5* mutant background, which prevents the cell death of P3.p and P4.p, the *ped-7* mutation results in ectopic vulval differentiation of these cells [1].

    A candidate gene approach had been instrumental in identifying several *P. pacificus* vulva mutations, such as the homologs of the *C. elegans* Hox genes *lin-39* and *mab-5* or of the *even-skipped* homolog, *vab-7* [24-26]. However, the identity of many interesting mutations cannot be guessed easily; moreover, candidate gene approaches bias against finding evolutionary novelties. Therefore, the Sommer laboratory undertook over the last few years the construction of genetic and physical maps for *Pristionchus pacificus*, as a prerequisite for mapping and identifying mutated loci without the bias of a candidate gene approach. The physical map was built from restriction fingerprinting of a BAC library [27]. The genetic map was constructed using DNA polymorphisms between divergent *P. pacificus* wild isolates. The choice was to focus on single-stranded conformation polymorphism (SSCP) in BAC ends,

thus directly providing an alignment of the physical and genetic maps [28]. The *P. pacificus* genome is going to be sequenced shortly (http://www.genome.gov/12511858).

*ped-7* is the first mutation to be identified using these physical and genetic maps. The *ped-7* locus was first mapped to a small chromosomal region using SSCP polymorphisms. Available sequence information from BAC-ends indicated the presence in this region of the homolog of *lin-17*, one of the *C. elegans* Wnt receptors. Molecular lesions in the coding region of this gene were then identified by sequencing six of the *ped-7* alleles [1], demonstrating that reduction-of-function mutations in this Wnt receptor cause the vulva phenotype. The *ped-7* locus was renamed *Ppa-lin-17*.

The involvement of several Wnt pathway components in vulval fate repression and in P7.p lineage polarity was further tested using morpholino injections in a sensitized hypomorphic *lin-17* mutant background. Inactivation of the Wnts *Ppa-lin-44* and *Ppa-egl-20* and of the dishevelled homolog *Ppa-mig-5* cause both gonad-independent vulva differentiation and P7.p lineage reversal, whereas inactivation of the APC homolog *Ppa-apr-1* and of the β-catenin homolog *Ppa-bar-1* only affect the former, and that of the divergent Ryk-like Wnt receptor *Ppa-lin-18* the latter [1]. A Wnt pathway is thus involved in repressing vulval fates in *P. pacificus* (Fig. 1).

**… whereas Wnt signaling promotes vulval fates in *C. elegans***

The repression of vulval fates by Wnt signaling in *Pristionchus pacificus* is a surprise, because Wnt pathway mutants show the opposite phenotype in *C. elegans*, namely a *decrease* rather than an increase in Pn.p vulval fate induction.

Recent results show that Wnt pathway components play a role in at least four cell specification events during *C. elegans* vulva development (Fig. 1):

1. Wnt signaling is first required for the maintenance of *lin-39*/Hox expression during the L2 stage, which prevents P(4-8).p (and sometimes P3.p) from fusing to the epidermal syncytium hyp7, and thus maintains their competence to adopt a vulval fate in the L3 stage [3-5]. Wnt signaling plays a major role at this step, but appears to be partially redundant with EGF signaling [3].

2. Wnt signaling plays a positive role in the adoption of vulval (1° and 2°) versus non-vulval (3°) fates by P(5-7).p in the L3 stage. The localized EGF signal from the anchor cell appears to play the major role in inducing vulval fates. Moroever, the role of Wnt signaling at this step has been more difficult to study because it is also required earlier in the L2 stage; however, careful observation of mutant phenotypes suggests that Wnt signaling also acts at this step [3,4,6], especially in some environmental conditions where it is partially functionally redundant with EGF-Ras signaling [7].

3. The Wnt pathway plays a role in P7.p lineage reversal, as in *P. pacificus* [12,29-31].

4. Wnt signaling plays a redundant role with the EGF pathway in the specification of P6.p inner granddaughter fates (repression of *zmp-1::GFP* transgene expression) [32].

Many Wnt pathway components come as a family of paralogs. The involvement of each paralog and the level of redundancy between them differs for each step. For example, the *lin-17*/Wnt receptor mutation causes a defect in P7.p lineage polarity in *C. elegans* as in *P. pacificus*, but does not directly affect the level of vulval cell fate induction in Pn.p cells [29], and does not result in gonad-independent vulval differentiation [1].

An apparent evolutionary inversion in the effect of Wnt signaling on vulval fate specification (step 2) has thus occurred between the evolutionary lineages leading to *C. elegans* and *P. pacificus* from their common ancestor (estimated to date back about 200 Myr ago). Analysis of additional species will be required to understand the origins of Wnt-based repression and activation of vulval fates, and whether there has been an actual switch in the sign of the effect, or whether each one was recruited de novo, independently. Particularly relevant features are the cellular source and target of Wnt signaling, i.e. the cell(s) expressing and receiving the Wnt signals, and the downstream effectors of Wnt signaling.

**A wiring change**

*What are the sources of Wnt signals and of negative signaling in the developing vulva?*

Little is currently known about the cells secreting Wnts around the vulva primordium and this promises to be an interesting topic in the near future.

In *C. elegans*, the putative Wnts and Wnt receptors are unknown for the two first Wnt-dependent steps, as is the signaling cell (no cell ablation studies giving us a hint); the receiving cells are the Pn.p cells, at least for step 1 [5]. Concerning steps 3 and 4, a *mom-2*/Wnt reporter gene is expressed in the anchor cell [12]; the latter could thus act as a spatial source for polarization of the P7.p and P6.p lineages.

In *P. pacificus*, the morpholino knockdown of two Wnts (the homologs of *lin-44* and *egl-20*) affects both vulval induction and P7.p polarity, but the source of these Wnts is unknown [1]. Strikingly however, ablations of P8.p mimic the gonad-independent vulval differentiation phenotype of the *Ppa-lin-17* mutant [33]. P8.p is thus a candidate as a source of Wnt signaling in the vulva repression step (and the gonad obviously is not). Alternatively, it could be relayed by the M cell (as is more clearly the case for a negative signaling process preventing P5.p and P7.p from adopting a 1° fate upon P6.p ablation) [33].

Negative Wnt signaling in *P. pacificus* appears to play a similar functional role as negative signaling through the synthetic Multivulva pathway in *C. elegans*, which appears to originate from the surrounding hyp7 epidermal syncytium [10,11]. In *C. elegans*, negative signaling affects the maintenance of *lin-39* expression in the Pn.p cells in the L2 and L3 stages (Fig. 1, steps 1 and 2)[34]. Since in these two steps this signaling results in the inhibited cell fusing to hyp7, it possibly acts through the expression in hyp7 and/or the Pn.p cells of general cell fusion-promoting molecules such as *eff-1* [35].

*Signal transduction*

In different systems, the inactivation of different Wnt pathway components sometimes show opposite effects relative to each other [36]: for example, in *C. elegans*, *apr-1*/APC RNAi inactivation has the same effect as the *bar-1*/β-catenin loss-of-function mutation in the L2 stage [5], but has an opposite effect in the L3 stage [6]. However, a conserved feature is the activations of Wnt receptors by Wnt signals. Therefore, the evidence presented by Zheng et al. using several *lin-17*/Wnt receptor mutant alleles demonstrates very convincingly that Wnt signals *repress* vulval fates in *P. pacificus*. Morpholino inactivation of three downstream components (see above) suggests that they are positively required for Wnt signal transduction. Especially, *apr-1*/APC and *bar-1*/β-catenin inactivations have the same effect, which is opposite to their effect at step 2 in *C. elegans*.

*Downstream effectors*

In *C. elegans*, LIN-39/Hox is a downstream transcriptional target of Wnt signaling in the Pn.p cells in steps 1-2 [3,6]. Wnt signaling positively requires the downstream transcription factor POP-1/TCF in step 2 [6] and regulates its nuclear accumulation in steps 3 and 4 [31]. In *P. pacificus*, the *lin-39* homolog is required to prevent cell death of P(5-8).p early on, but is not positively required for P(5-7).p cell fate differentiation [37], and the transcription factors downstream of Wnt signaling are so far unknown.

Much remains thus to be explored in these systems. Hopefully, further studies will unravel the detailed molecular changes that underlie this dramatic change in Wnt pathway recruitment, either by cis-regulatory evolution affecting Wnt expression or transcriptional target regulation, or through protein evolution that may transform a positive regulation into a negative one (or conversely).

**Pathway redundancy and evolutionary flexibility**

Evolution in the recruitment of signaling pathways thus occurs in nematode vulva development, a system which, like many others, uses redundant pathways and mechanisms for cell specification [38]. The use of multiple mechanisms for cell fate patterning ensures a robust outcome, even when the developmental system is faced with stochastic noise or environmental variations. As a consequence, the system may also be robust to a variety of genetic changes, and this may allow for the evolution of underlying mechanisms [14,39,40]. Very interestingly, the inhibitory role of P8.p and the requirement for positive signaling from the gonad vary within the genus *Pristionchus* and between *P. pacificus* isolates [18]. It is possible that the weight of negative signaling through the Wnt pathway evolves at this microevolutionary scale.

**Acknowledgements.** I am very grateful to R. Sommer for discussions and to C. Braendle for helpful comments on the manuscript. Work in the laboratory is supported by the CNRS, the ARC, a Developmental Biology ACI grant from the Ministry of Research (France) and the HFSPO.

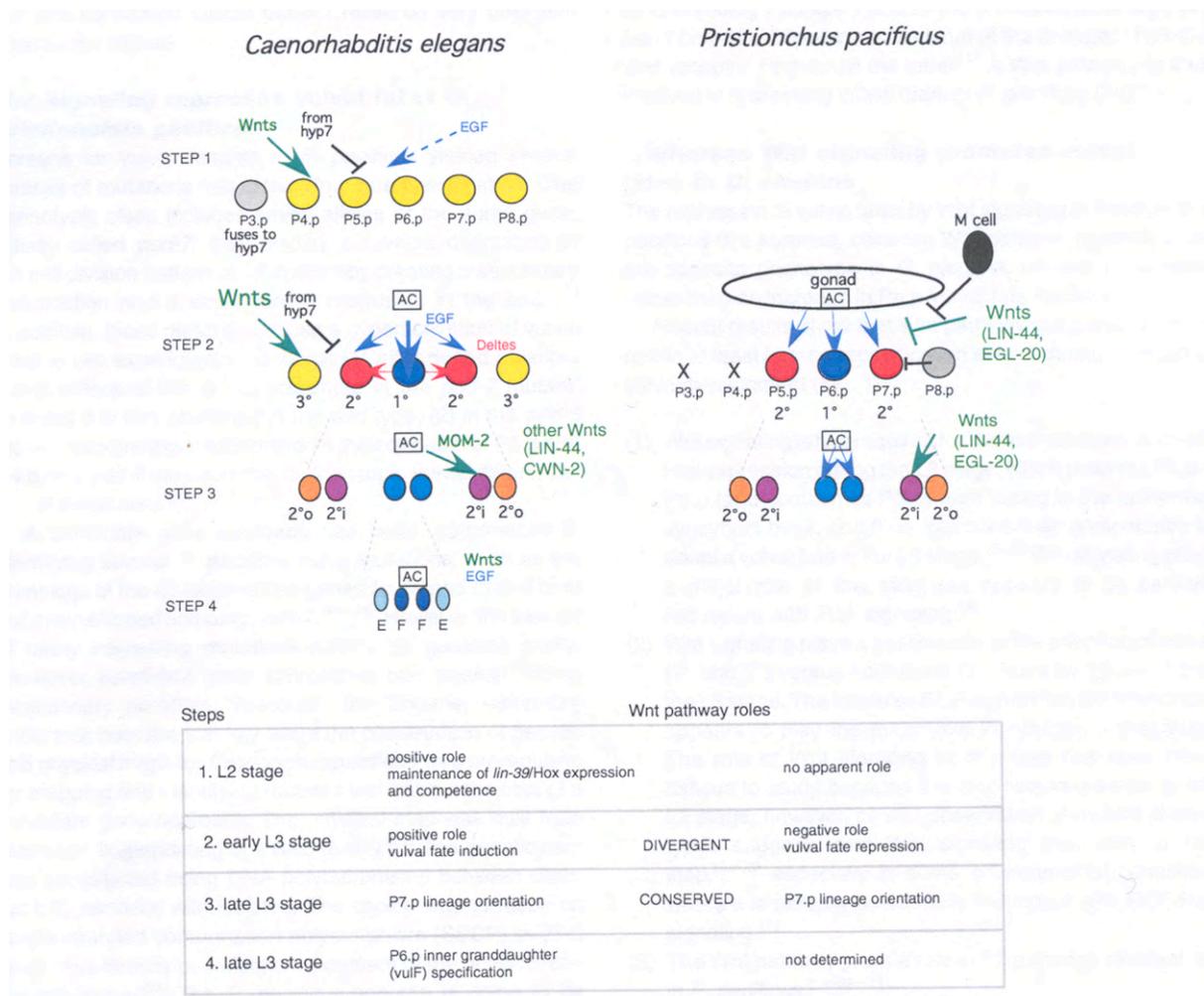

**Figure legend**
**Intercellular signaling in *C. elegans* and *P. pacificus* vulva development.**
The four steps where Wnt signaling has been shown to play a role in *C. elegans* are indicated. See text for references. The evolutionary change in the role of Wnt signaling between *C. elegans* and *P. pacificus* is seen at step 2, i.e. the specification of vulval versus non-vulval Pn.p cell fates. Dashes indicate that the corresponding intercellular signal is not the major source of fate specification, but acts redundantly. The source of Wnt signals affecting steps 1 and 2 is not known. The receiving cells for Wnt and negative signaling at steps 1 and 2 are P(3-8).p in *C. elegans*, and are unknown in *P. pacificus*. There are five Wnts in *C. elegans*: MOM-2, LIN-44, EGL-20, CWN-1 and CWN-2. The cell fates and molecular pathways are independently color-coded. Grey: non-competent Pn.p cell. AC: anchor cell. 2°i: inner 2° fate. 2°o: outer 2° fate.